\documentclass{sig-alternate-05-2015}

\usepackage{balance}
\usepackage{float}
\usepackage{subcaption}
\usepackage{hyperref}

\graphicspath{{./figures/}}

\begin{document}

\CopyrightYear{2016}
\setcopyright{acmlicensed}
\conferenceinfo{MM '16,}{October 15 - 19, 2016, Amsterdam, Netherlands}
\isbn{978-1-4503-3603-1/16/10}\acmPrice{\$15.00}
\doi{http://dx.doi.org/10.1145/2964284.2973801}

\clubpenalty=10000
\widowpenalty = 10000

%

\title{CNNdroid: GPU-Accelerated Execution of Trained\\Deep Convolutional Neural Networks on Android}
%
%
%
%
%
%

\numberofauthors{1} 
%
\author{
%
%
\alignauthor
Seyyed Salar Latifi Oskouei, Hossein Golestani, Matin Hashemi\titlenote{\small Contact author: Matin Hashemi, matin@sharif.edu}\\
       \affaddr{Sharif University of Technology}\\
       \email{salarlatifi@ee.sharif.edu, hossein\_golestani@ee.sharif.edu, matin@sharif.edu}
\and  
\alignauthor Soheil Ghiasi\\
       \affaddr{University of California, Davis}\\
       \email{ghiasi@ucdavis.edu}
}

\maketitle
\begin{abstract}
Many mobile applications running on smartphones and wearable devices would potentially benefit from the accuracy and scalability of deep CNN-based machine learning algorithms. However, performance and energy consumption limitations make the execution of such computationally intensive algorithms on mobile devices prohibitive. We present a GPU-accelerated library, dubbed CNNdroid \cite{CNNdroid}, for execution of trained deep CNNs on Android-based mobile devices. Empirical evaluations show that CNNdroid achieves up to $60$X speedup and $130$X energy saving on current mobile devices. The CNNdroid open source library is available for download at \url{https://github.com/ENCP/CNNdroid} 
\end{abstract}

%
%

\ccsdesc[500]{Computing methodologies~Neural networks}
\ccsdesc[500]{Human-centered computing~Open source software}
\ccsdesc[500]{Human-centered computing~Ubiquitous and mobile devices}
\ccsdesc[500]{Computer systems organization~Embedded systems}
\ccsdesc[500]{Computer systems organization~Parallel architectures}

%
%

%
%


\keywords{Deep Learning, Deep Convolutional Neural Network (CNN), Mobile GPU, Performance Optimization, Low Energy Consumption, Open Source Software, Android, RenderScript}

\section{Introduction}
\label{sec:intro}

\begin{figure}
\centering
\includegraphics[width=3.3in]{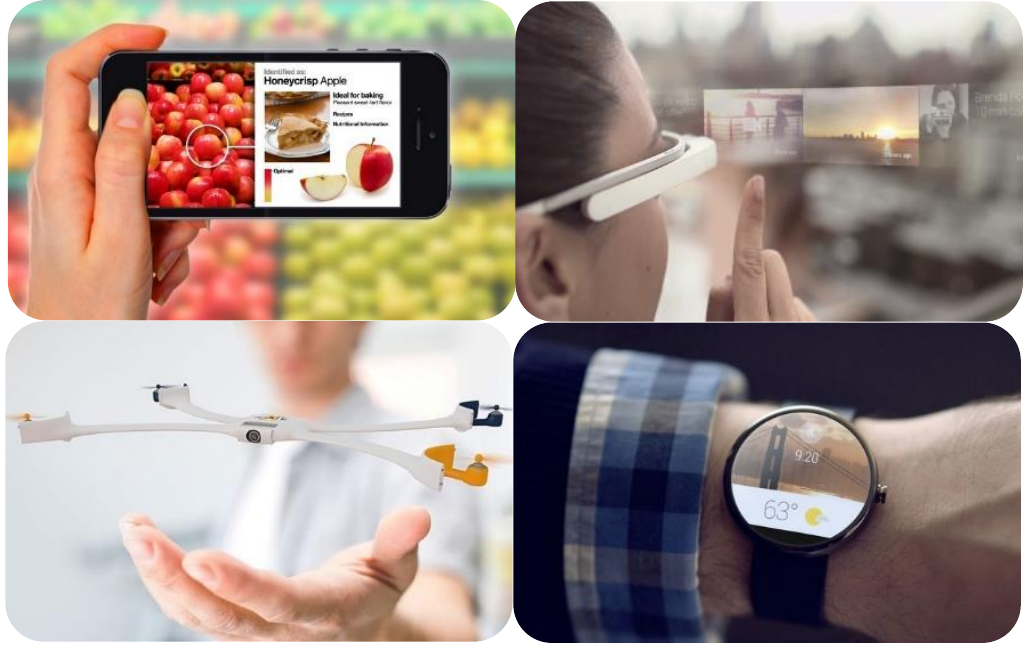}
\caption{\small Example applications of deep CNNs in mobile systems. Image credits: IBM Research, Guardianlv, Nixie, Android Wear.} 
\label{fig:application}
\end{figure}

Mobile platforms such as smartphones, wearable devices, tiny autonomous robots and IoT devices have been increasingly finding their way into many areas (Figure \ref{fig:application}). Numerous applications, such as speech recognition and image recognition \cite{mobile1}, would potentially benefit from local execution of accurate machine learning algorithms on mobile devices. Local execution allows data to stay on the mobile device and hence avoids latency issues of cloud-assisted processing.

Deep CNNs can achieve state-of-the-art results in terms of both prediction accuracy and scalability. However, they are highly computationally intensive, and hence, not practical on current mobile devices without acceleration. 

Many hardware-based solutions have been proposed for acceleration of deep CNNs \cite{isscc_2016_eyeriss,ghiasi_2016}. IBM has also introduced a neuromorphic CMOS chip for execution of learning applications on smartphones and IoT devices \cite{truenorth}. While promising, such solutions are still in early stages of development and not available on current mobile devices.

As opposed to hardware-based engines, GPU already exists in many current mobile devices and can be programmed completely in software. Therefore, parallel processing capabilities of mobile GPUs can be exploited to accelerate deep CNN computations on \emph{current} mobile devices. 

On server and desktop platforms, there exists many GPU-accelerated deep CNN libraries \cite{caffe,torch,theano2010,tensorflow,cuDNN,cu-con,velesnet}. However, because of architecture  differences (Section \ref{sec:background:arch}), mere porting of such libraries to mobile platforms yields sub-optimal performance or is impossible in some cases (Section \ref{sec:background:desktop}). 

On mobile platforms, to the best of our knowledge, such GPU-accelerated libraries are not available. The few existing mobile libraries for CNN computations \cite{caffe_mobile, torch_mobile, awesomecnn_mobile, f_mobile} are limited to the processing power of multi-core mobile CPUs (Section \ref{sec:background:mobile}). 

We present an open source GPU-accelerated library, dubbed CNNdroid, which is specifically designed and optimized for execution of trained deep CNNs on Android-based mobile devices. The main highlights of CNNdroid are as follows.

\newpage
\begin{enumerate}
\item Support for nearly all CNN layer types (Section \ref{sec:library:layers}).
\item Compatible with CNN models trained by common desktop/server libraries, namely, Caffe \cite{caffe}, Torch \cite{torch} and Theano \cite{theano2010} (Section \ref{sec:library:script}). 
\item Easy to configure and integrate into any Android app without additional software requirements (Section \ref{sec:library:test}).
\item User-specified maximum memory usage (Section \ref{sec:library:memory}).
\item GPU or CPU acceleration of supported CNN layers (Section \ref{sec:library:acceleration}).
\item Automatic tuning of performance (Section \ref{sec:library:tuning}).
\item Up to $60$X speedup and up to $130$X energy saving on current mobile devices (Section \ref{sec:Results}).
\end{enumerate}

\section{Background and Related Work}
\label{sec:Background}

\subsection{Comparing Mobile and Desktop GPUs}
\label{sec:background:arch}

A modern graphics processing unit (GPU), in addition to computer graphics, can be programmed for general purpose computations as well. While desktop GPUs have long been programmable, 
major mobile chip manufacturers have recently made the GPU hardware available for general purpose computations. 
Due to strict area and power constraints, mobile GPUs have important differences  with their desktop counterparts. 

\begin{figure}
\centering
\includegraphics[width=3.3in]{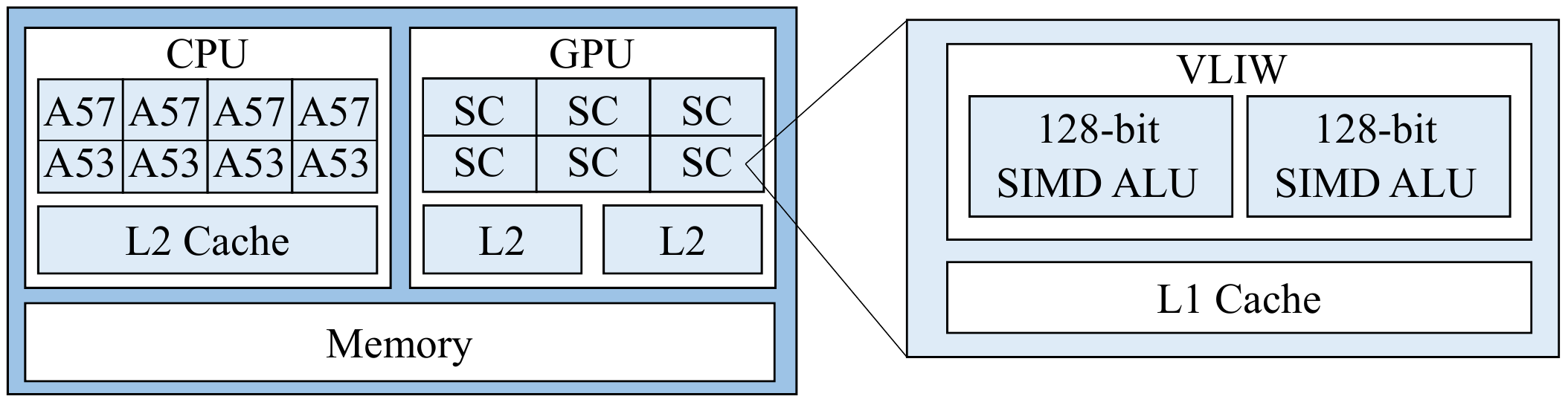}
\caption{\small Example: Exynos 5433 mobile processor with ARM A53 / A57  CPU and Mali T-760  GPU (SC: Shader Core, VLIW: Very Long Instruction Word, SIMD: Single Instruction Multiple Data).}
\label{fig:Midgard}
\end{figure}

A modern mobile GPU is typically composed of several programmable parallel computing units called Shader Cores (SC). Every shader core is composed of several parallel ALUs. 
For example, Samsung Exynos 5433 chip is composed of ARM A53/A57 CPU and Mali T-760 GPU (Figure \ref{fig:Midgard}). Each SC in T-760 GPU has two 128-bit ALUs in VLIW format. Each 128-bit ALU is capable of performing SIMD operations, i.e., two 64-bit, four 32-bit or eight 16-bit operations in parallel \cite{Mali_OpenCL}. 
In comparison with desktop GPUs, the above parallel ALU architecture relies more on software and compiler than dynamic hardware scheduler in efficient execution of parallel threads. 

More importantly, fast shared memory in thread blocks which are present in desktop GPUs and widely employed in many CUDA-based desktop libraries are not available in mobile GPUs. 

There are some differences on the software side as well. For example in RenderScript, Android's parallel computing platform \cite{RenderScript}, thread synchronization is not available. In addition, there must be a one-to-one correspondence between parallel threads and the data items inside one of the memory buffers that parallel threads work on. 

\subsection{Comparing CNNdroid with Desktop\\Libraries}
\label{sec:background:desktop}

On server and desktop platforms, there exists many libraries such as Caffe \cite{caffe}, Torch \cite{torch}, Theano \cite{theano2010}, TensorFlow \cite{tensorflow}, cuDNN \cite{cuDNN}, cuda-convnet \cite{cu-con}, and Velesnet \cite{velesnet}, which employ GPU-based parallel processing for acceleration of deep CNN computations. 
However, the acceleration methodologies and parallel algorithms of such libraries could not be directly utilized in mobile platforms due to the existing hardware and software differences.

For example in Caffe \cite{caffe}, the convolution operation is unrolled and converted to matrix multiplication, which requires considerable amount of memory and therefore is not suitable for mobile devices with small cache and memory sizes. 
As another example, the parallel algorithm in Theano \cite{theano2010} is similar to CNNdroid but without efficient use of SIMD units in mobile GPUs. Refer to Section \ref{sec:library:acceleration} for details.

More importantly, desktop libraries take advantage of thread management facilities provided by desktop GPUs and CUDA framework, such as fast shared memory and thread synchronization, which are not available in mobile GPUs and RenderScript.

\subsection{Comparing CNNdroid with Mobile\\Libraries}
\label{sec:background:mobile}

On mobile platforms, to the best of our knowledge, only few deep CNN libraries exist \cite{caffe_mobile, torch_mobile, awesomecnn_mobile, f_mobile}. All such libraries, including Caffe Mobile \cite{caffe_mobile} and Torch Mobile \cite{torch_mobile}, are limited to the processing power of multi-core mobile CPUs, while CNNdroid efficiently employs both GPU and CPU (Section \ref{sec:library:acceleration}). 

In addition, CNNdroid is compatible with CNN models trained by Caffe \cite{caffe}, Torch \cite{torch} and Theano \cite{theano2010}, which facilitates the process of porting the trained models to mobile devices (Section \ref{sec:library:script}).

The existing libraries require installation of Android NDK alongside Android SDK, while in CNNdroid, only Android SDK is required.

\section{CNNdroid Library}
\label{sec:library}

\subsection{CNN Layer Types}
\label{sec:library:layers}

CNNdroid library supports nearly all common types of CNN layers, namely, convolution, max/mean pooling, fully connected, rectified linear unit, local response normalization and softmax. Detailed description of every layer type and its corresponding parameters are available in the library documentations \cite{CNNdroid}. Other new layers may be added as well, due to the open source nature of the library.

\subsection{Model Preparation}
\label{sec:library:model}
\label{sec:library:script}

\textbf{Model Conversion Scripts:} Figure \ref{fig:deploy} shows an overview of the steps involved in deploying trained CNN models on mobile devices. CNNdroid library provides a set of scripts which take the models trained by common desktop/server libraries, namely, Caffe \cite{caffe}, Torch \cite{torch} and Theano \cite{theano2010}, as input and convert them into CNNdroid format. Therefore, the models which are trained by these libraries can be executed by CNNdroid library on mobile devices.

It is possible to write similar scripts for other libraries as well. CNNdroid uses MessagePack serialization format \cite{msgpack} for storing layer parameters in the trained model. Detailed procedure is presented in the library documentations \cite{CNNdroid}.

\textbf{NetFile:} The developer needs to prepare a $.txt$ file, called $NetFile$, similar to the $.prototxt$ file in Caffe \cite{caffe}. The $NetFile$ specifies layer setup of the trained model, i.e., order of the CNN layers along with their configurations, e.g., padding and stride values of convolution layers. Figure \ref{fig:NetFile} presents an example. Detailed instructions for preparing the $NetFile$ as well as a few examples are included in the library documentations \cite{CNNdroid}.

The $NetFile$ also specifies three configuration parameters (Figure \ref{fig:NetFile}), namely, \texttt{allocated\_ram} which specifies the maximum amount of memory that CNNdroid engine is allowed to allocate at runtime (Section \ref{sec:library:memory}), \texttt{execution\_mode} which selects between sequential or parallel execution modes of the library (Section \ref{sec:library:acceleration}), and \texttt{auto\_tuning} which specifies whether or not auto-tuning is turned on (Section \ref{sec:library:tuning}). 

\subsection{Model Execution}
\label{sec:library:test}

\begin{figure}
\centering
\includegraphics[width = 3.3 in, keepaspectratio]{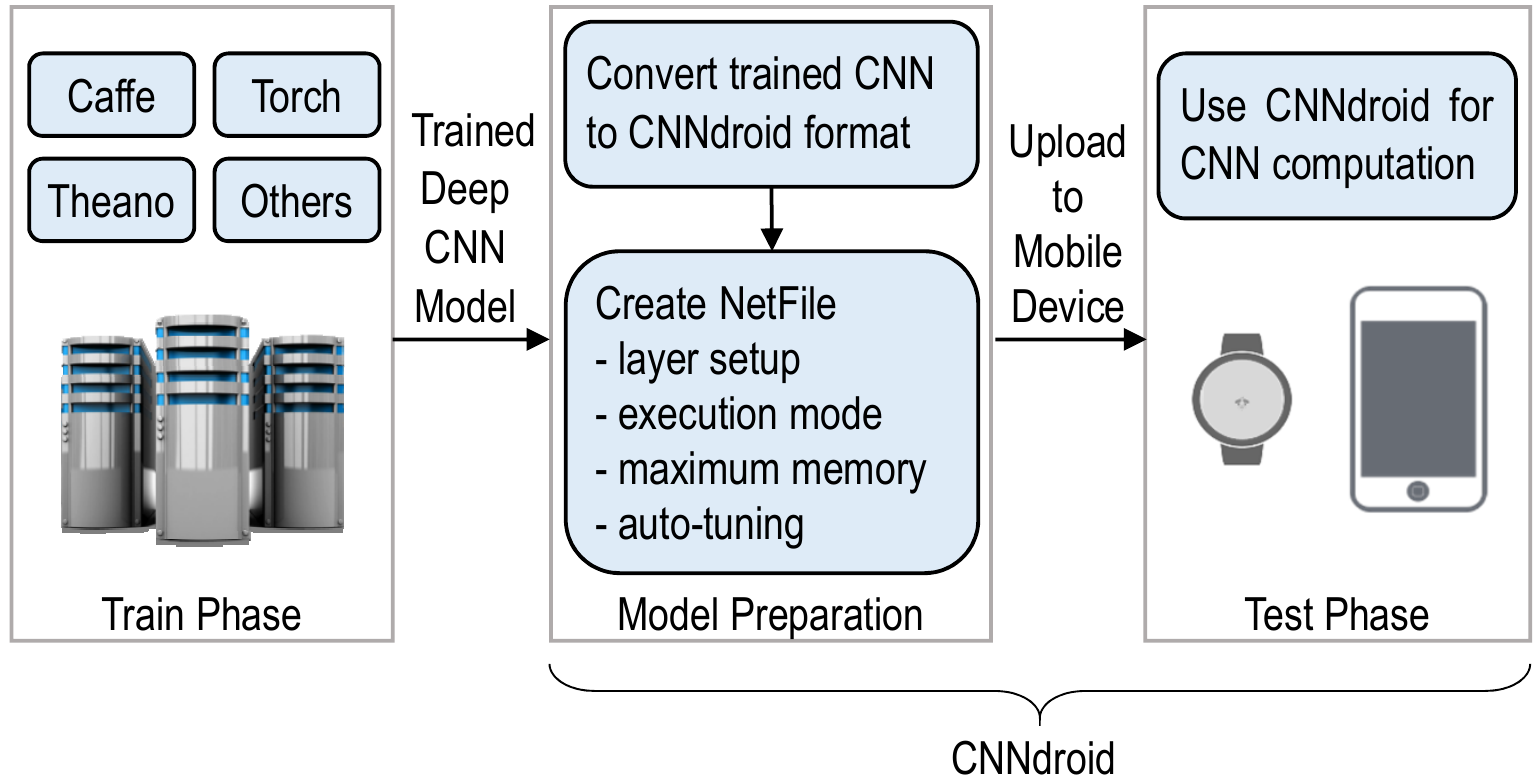}
\vskip -1mm
\caption{\small Overview of CNNdroid's model deployment procedure.}
\label{fig:deploy}
\end{figure}

Once the trained model and its corresponding $NetFile$ are both uploaded to mobile device (Figure \ref{fig:deploy}), the model can be executed in the target Android application in a few simple steps as described below (Figure \ref{fig:running}).

The first step is to include the provided CNNdroid library files. Note that CNNdroid library is self-sufficient and does not require installation of third-party libraries. In addition, it does not require installation of Android NDK alongside Android SDK. 

Next, \texttt{RenderScript} and \texttt{CNNdroid} objects are constructed (Figure \ref{fig:running}, steps $2$ and $3$). The CNNdroid constructor takes the provided $NetFile$ as input and automatically creates the corresponding objects for the network layers. 

Finally, \texttt{compute} function of the constructed \texttt{CNNdroid} object is called, which automatically executes the trained model on either a single image or a batch of images. 

\subsection{Memory Allocation}
\label{sec:library:memory}

\begin{figure}
\centering
\includegraphics[width = 3.3 in, keepaspectratio]{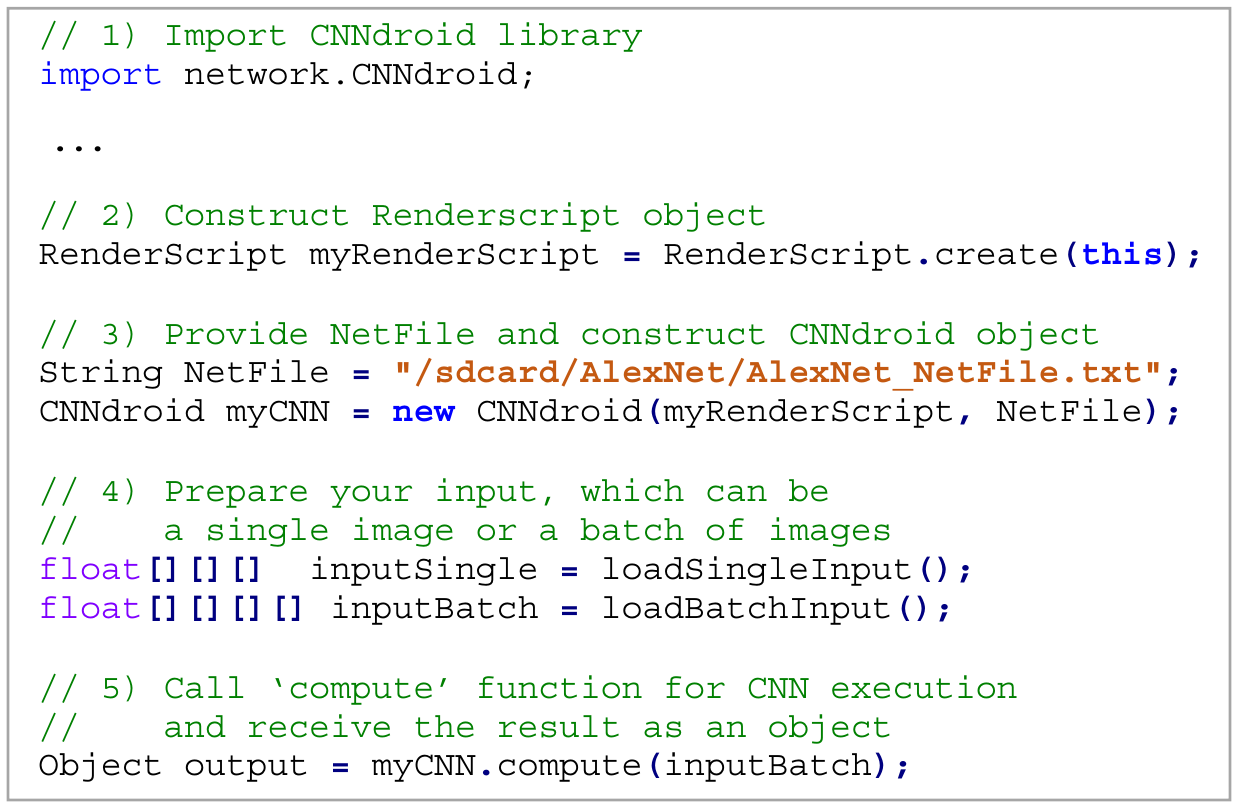}
\vskip -1mm
\caption{\small Simple steps involved in using CNNdroid for execution of trained deep CNN models on Android apps. Refer to the library documentation for up-to-date details and sample projects \cite{CNNdroid}.} 
\label{fig:running}
\end{figure}

The trained CNN model, which is uploaded to SD card of the mobile device, contains layer parameters in form of matrices. Inside the \texttt{compute} function (Figure \ref{fig:running}, step $5$), and before execution of every layer, the corresponding matrices are automatically loaded from SD card into RAM, which incurs an overhead.

In order to reduce this overhead, CNNdroid selects certain layers and keeps their data in RAM, while other layers are allocated and de-allocated every time. 
The selection procedure is automatically done in \texttt{CNNdroid} constructor (Figure \ref{fig:running}, step $3$). Starting from the largest layer, as many layers as possible are selected, till the sum of their memory sizes reaches the maximum developer-specified amount, i.e., the \texttt{allocated\_ram} parameter in the $NetFile$.

Note that the \texttt{allocated\_ram} parameter cannot be arbitrarily large due to practical limitations. For example, Android $5$ limits the memory usage of every app to $512$MB. 

\begin{figure}[p]
\centering
\includegraphics[height = 3.7 in, keepaspectratio]{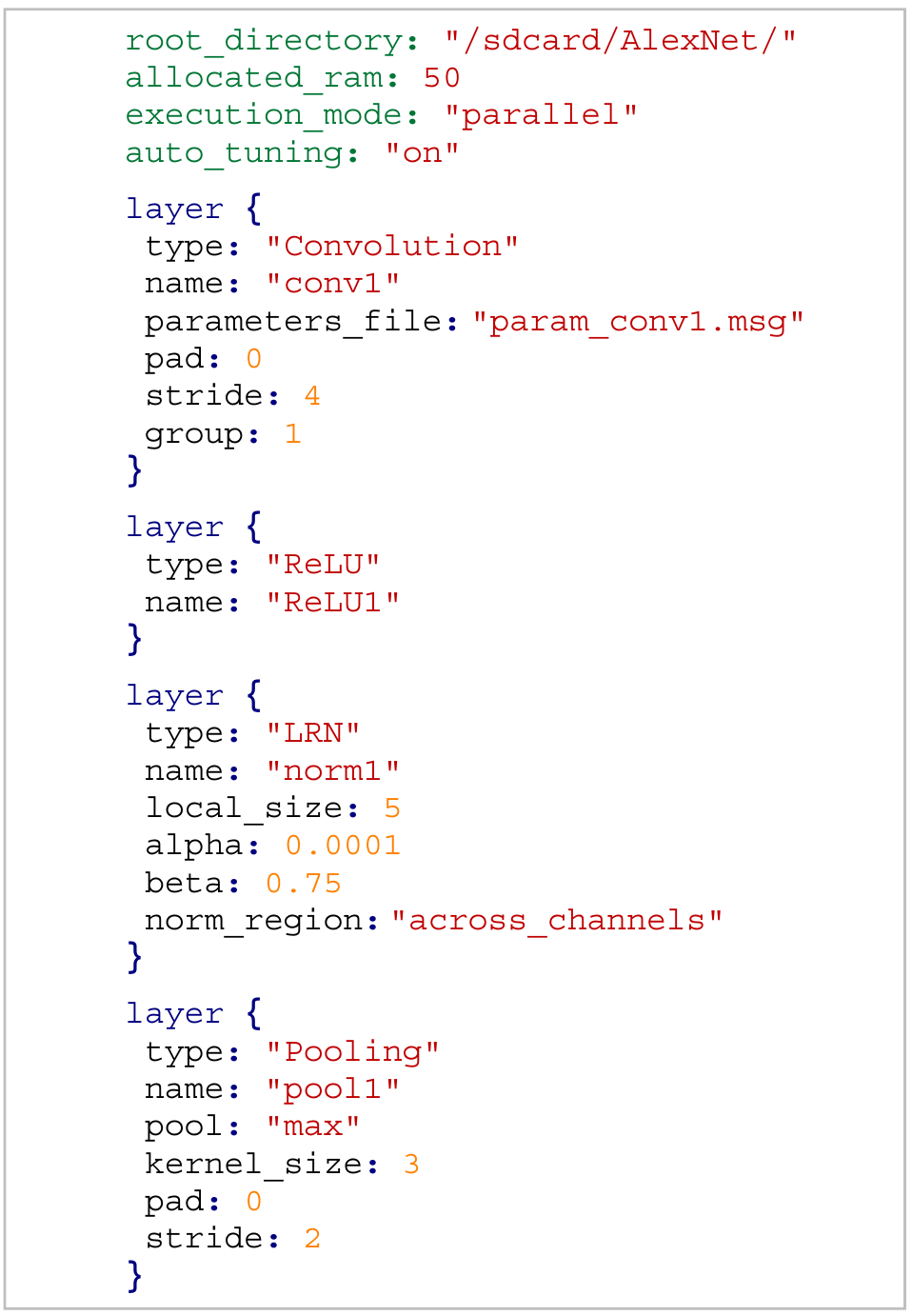} 
\vskip -1mm
\caption{\small Example NetFile showing three layers of AlexNet \cite{ImageNet}, along with \texttt{allocated\_ram}, \texttt{execution\_mode} and \texttt{auto\_tuning} parameters.}
\label{fig:NetFile}
\end{figure}

\subsection{Acceleration Methods}
\label{sec:library:acceleration}

Different methods are employed in acceleration of different layers in CNNdroid.
Convolution and fully connected layers, which are data-parallel and normally more compute intensive, are accelerated on the mobile GPU using RenderScript framework. 

A considerable portion of these two layers can be expressed as dot products. In specific, in the convolution layer, kernels get convoluted with the input frames, and in the fully connected layer, the computation can be expressed as a matrix to vector multiplication. The dot products are more efficiently calculated on SIMD units of the target mobile GPU. Therefore, we divide the computation in many vector operations and use the pre-defined $dot$ function of the RenderScript framework. In other words, we explicitly express this level of parallelism in software, and as opposed to CUDA-based desktop libraries, do not leave it to GPU's hardware scheduler.

Comparing with convolution and fully connected layers, other layers are relatively less compute intensive and not efficient on mobile GPU. Therefore, they are accelerated on multi-core mobile CPU via multi-threading. 
Since ReLU layer usually appears after a convolution or fully connected layer, it is embedded into its previous layer in order to increase the performance in cases where multiple images are fed to the CNNdroid engine.

In addition to above parallel implementations, CNNdroid also includes sequential (single-thread) implementations of all layers. The execution will be sequential or parallel depending on the \texttt{execution\_mode} parameter specified in the $NetFile$ (Figure \ref{fig:NetFile}).

\subsection{Auto-Tuning}
\label{sec:library:tuning}

In order to reach better performance across different Android based mobile devices, our GPU-accelerated parallel algorithms are developed with tuning parameters which select the amount of work assigned to parallel GPU threads and the amount of work assigned to SIMD ALUs in execution of every GPU thread. The tuning parameters basically determine the granularity of parallelism.

If turned on in the $NetFile$ (Figure \ref{fig:NetFile}), the auto-tuner is automatically executed when the Android app is launched for the first time. It executes the CNN model for a number of predefined scenarios on the mobile device, measures their runtime and saves the optimum tuning parameters for future executions. As a result, the first launch of the application takes much longer time. 
For the purpose of clear and fair comparisons, auto-tuning is turned off in our experiments in Section \ref{sec:Results}.

\section{Empirical Evaluation}
\label{sec:Results}

CNNdroid is empirically evaluated on two mobile devices, namely, Samsung Galaxy Note $4$ and HTC One M$9$.
We employ three well-known benchmark CNNs, namely, LeNet network for MNIST dataset \cite{MNIST}, Alex Krizhevsky's network for CIFAR-10 (Alex's CIFAR-10) \cite{CIFAR} and Alex Krizhevsky's network for ImageNet 2012 dataset (AlexNet) \cite{ImageNet}. 

\begin{figure}
	\centering
	\includegraphics[width = 2.6 in]{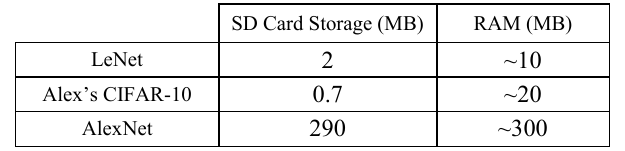}
	\vskip -2mm	
	\caption{\small Storage and memory requirements of benchmark CNNs when ported to CNNdroid format.}
		\vskip -2mm	
	\label{tbl:memory}
\end{figure}

Layer setup of the benchmark CNNs are shown in Figure \ref{tbl:test_net}. We also measured the storage and memory requirement of the benchmark CNNs when ported to CNNdroid format. The results are reported in Figure \ref{tbl:memory}.

We execute forward paths of the benchmark CNNs on the mobile devices and measure their accuracy, runtime and energy consumption. 
All benchmarks accept batches of $16$ images as input in our experiments.
Before running every experiment, mobile devices are fully charged and put into airplane mode and minimum screen brightness. 
The measurements reported below are only for CNN execution and not for loading the network parameters from SD card, because in our benchmarks network parameters are loaded only once in the beginning but CNN execution is performed on every input image.

\begin{figure}
\scriptsize
\center{ \begin{tabular}{|c|c|c|c|}
\hline
Layer & LeNet  & Alex's CIFAR-10  & AlexNet \\
\hline 

1	& Conv & Conv & Conv+ReLU \\
2	& Pooling & Pooling+ReLU & LRN \\
3	& Conv & Conv+ReLU & Pooling \\
4	& Pooling & Pooling & Conv+ReLU \\
5	& FC+ReLU & Conv+ReLU & LRN \\
6	& FC & Pooling & Pooling \\
7	& - & FC & Conv+ReLU \\
8	& - & FC & Conv+ReLU \\
9	& - & - & Conv+ReLU \\
10	& - & - & Pooling \\
11	& - & - & FC+ReLU \\
12	& - & - & FC+ReLU \\
13	& - & - & FC \\
\hline
\end{tabular}
}
\vskip -2mm
\caption{\small Layer setup of benchmark CNNs.}
\label{tbl:test_net}
\end{figure}

\subsection{Accuracy}

In order to measure CNNdroid accuracy, output of the last network layer in both CNNdroid and Caffe are compared for the same input. The resulting mean square error is in the order of $10^{-12}$, which means there is no meaningful difference and CNNdroid is correctly implemented.

\subsection{Performance}
\label{sec:Performance}

\begin{figure}
	\centering
	\includegraphics[width = 3.3 in]{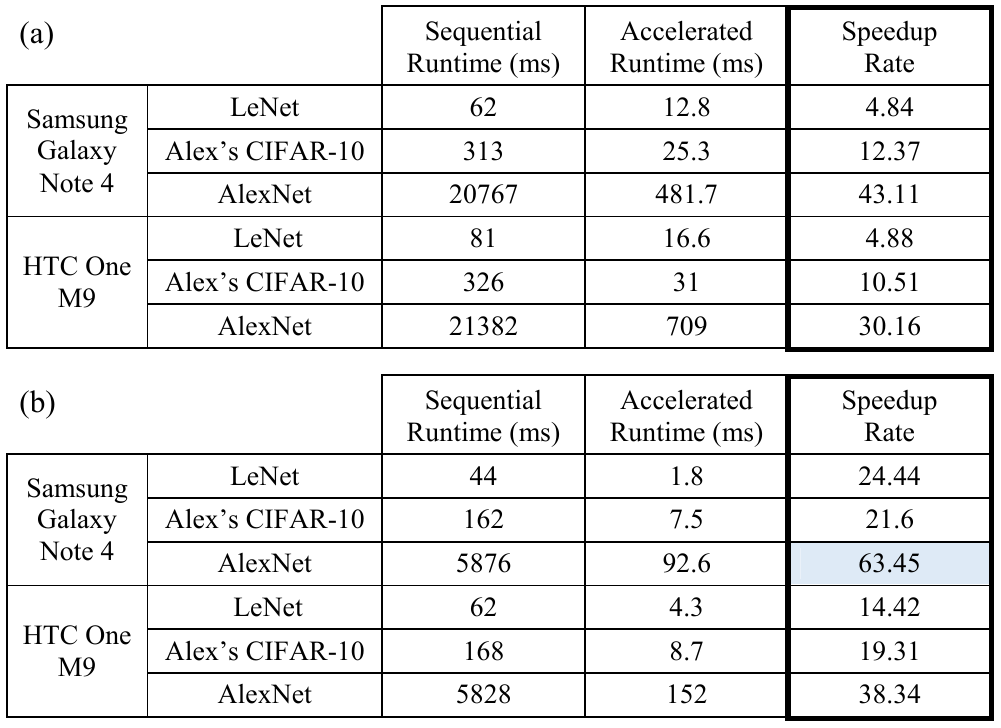}
	\caption{\small Average runtime of (a) the entire CNN and (b) the heaviest convolution layer, per image in a batch of $16$ images, and the corresponding speedup rate.}
	\label{tbl:speedup}
\end{figure}

Figure \ref{tbl:speedup}.a shows the total measured runtime for CPU-only sequential CNN implementation as well as the speedup gained by GPU acceleration. The reported values are the average of ten executions. 

Note that real-time performance is achieved on mobile devices in LeNet and Alex's CIFAR-10 benchmarks. For instance on the HTC One M9 device, the accelerated implementation achieved $60.2$ and $32.2$ frames per second for LeNet and Alex's CIFAR-10 benchmarks, respectively.

We also measured runtime of the heaviest convolution layer in order to observe direct impact of the GPU-based acceleration (Figure \ref{tbl:speedup}.b).
The highest achieved speedup is $63.4$X for AlexNet benchmark on Galaxy Note $4$ device with Mali-T760 GPU. This GPU has $6$ shader cores, each with two $128$-bit ALUs. Since all elements of the matrices in our CNN model are 32-bit floating point values, a maximum of $6 \times 2 \times \frac{128}{32} = 48$ operations may run in parallel. In other words, the maximum theoretically achievable speedup is $48$X. Therefore, the measured $63.4$X speedup most probably comes from other factors such as software language performance of RenderScript(c99) over Java or cache effects.

Note that the benchmark CNNs in our experiments accept batches of $16$ images as input, and the runtime values reported in Figure \ref{tbl:speedup} are per image. It is recommended to process a batch of input images rather than a single image to get  higher performance in CNNdroid.

As for the performance comparison of our mobile devices, we see that the overall speedup in AlexNet, which is a large network, is approximately $30\%$ higher on Galaxy Note $4$ compared with HTC One M9. This can be either the result of lower GPU frequency of HTC One M9 or its aggressive throttling policy in order to prevent overheating issues in long runtimes. 

\subsection{Energy Consumption}

We measured power and energy consumption per image for AlexNet benchmark on HTC One M9 by employing ``Qualcomm Trepn Profiler'' application \cite{Trepn}.

The GPU accelerated execution consumes around $523~mW$ power and $0.4~J$ energy while the CPU-only sequential execution consumes $2338~mW$ power and $51.6~J$ energy. As a result, the GPU accelerated execution consumes $51.6 \div 0.4 = 129$X less battery energy.

It should be noted that we observed about $20\%$ variability in our measurements which is expected since Trepn Profiler provides a software-only method for measuring energy consumption of a single app.


\section{Conclusions}
\label{sec:Conclusion}

We introduced CNNdroid, an open source GPU accelerated deep CNN library for Android-based mobile devices. Empirical evaluations demonstrated up to $60$X speedup and up to $130$X energy saving. The source code, documentation and sample projects are published online \cite{CNNdroid}. 



%
\bibliographystyle{unsrt} 
\bibliography{sigproc}  
\balance 
\end{document}